\documentstyle[draft,aps,epsfig]{revtex}
\begin{document}
\draft
\title{Problems With Extracting $m_s$ from Flavor Breaking in Hadronic 
$\tau$ Decays  }
\author{Kim Maltman\thanks{e-mail: maltman@fewbody.phys.yorku.ca}}
\address{Department of Mathematics and Statistics, York University, \\
          4700 Keele St., Toronto, Ontario, CANADA M3J 1P3 \\ and}
\address{Special Research Center for the Subatomic Structure of Matter, \\
          University of Adelaide, Australia 5005}
\maketitle
\begin{abstract}
A numerical error is pointed out in the existing expression for
the ${\cal O}(\alpha_s^2)$ 
longitudinal component of the squared-mass
($D=2$) contribution to the hadronic $\tau$ decay rate.
The corrected version is found to be such that, to ${\cal O}(\alpha_s^2)$,
each term in the resulting series is larger than the previous one,
hence ruling out the direct use of
flavor breaking in hadronic $\tau$ decays as
a means of extracting
the strange quark mass.  An alternate approach, in which one
uses the model spectral functions previously
developed for the strangeness-changing
scalar channel (and employed in
alternate sum rule analyses of $m_s$) as input to 
the $\tau$ decay analysis, is shown to
provide mutual consistency checks for the two
different methods.  The results for $m_s$ implied by $\tau$
decay data, and the existing model spectral function are
also presented.
\end{abstract}
\def\mstitle{$m_s$}
\section{Introduction}
As is well-known, the inclusive hadronic $\tau$ decay ratio
\begin{equation}
R_\tau \equiv \frac{ \Gamma [\tau^- \rightarrow \nu_\tau
\, {\rm hadrons}\, (\gamma)]}{ \Gamma [\tau^- \rightarrow
\nu_\tau e^- {\bar \nu}_e (\gamma)] } ,
\label{one}\end{equation}
(where $(\gamma )$ indicates additional photons or lepton pairs)
may be rather reliably computed using techniques based on the OPE
and perturbative QCD 
(pQCD)\cite{tau1,tau2,tau3,tau4,tau5,tau6,tau7,tau8,tau9,tau10}.
(A very clear recent review of the
theoretical situation is given in Ref.~\cite{pich97}.)
One begins with the representation of this ratio in terms of
hadronic spectral functions,
\begin{equation}
R_\tau   =  
12 \pi S_{EW}\, \int^{m_\tau^2}_0 {\frac{ds} {m_\tau^2 }} \,
\left( 1-{\frac{s}{m_\tau^2}}\right)^2 
\left[ \left( 1 + 2 {\frac{s}{m_\tau^2}}\right) 
\, {\rm Im}\, \Pi^{(1)}(s)  + \, {\rm Im}\, \Pi^{(0)}(s) \right] \, 
\label{hadronicrep}\end{equation}
where $S_{EW}=1.0194$ represents the leading electroweak 
corrections\cite{ms88},
$J=0,1$ labels the hadronic rest-frame angular momentum,
and the $\Pi^{(J)}(q^2)$ are given in terms of the vector and axial
vector current spectral functions $\Pi^{(J)}_{ij;V,A}(q^2)$ (with
$ij=ud,us$) by
\begin{equation}
\Pi^{(J)}(s)  \equiv
|V_{ud}|^2 \, \left[ \Pi^{(J)}_{ud,V}(s) + \Pi^{(J)}_{ud,A}(s) \right]
+ |V_{us}|^2 \, \left[ \Pi^{(J)}_{us,V}(s) + \Pi^{(J)}_{us,A}(s) \right]
,
\end{equation}
where 
\begin{equation}
i \int d^4x \, e^{iqx} 
\langle 0|T(J^{\mu}_{ij;V,A}(x) J^{\nu}_{ij;V,A}(0)^\dagger)|0\rangle \, 
\equiv (-g^{\mu\nu} q^2 + q^{\mu} q^{\nu}) \, \Pi_{ij;V,A}^{(1)}(q^2)
 + q^{\mu} q^{\nu} \, \Pi_{ij;V,A}^{(0)}(q^2)\, ,
\end{equation}
with $J^{\mu}_{ij;V,A}$ the standard vector and axial currents 
involving flavors $ij$, defines $\Pi^{(J)}_{ij;V,A}(q^2)$.
Using the analyticity properties of the correlators, the
hadronic representation, Eq.~(\ref{hadronicrep}), can then be converted into
a contour integral representation
\begin{equation}
R_\tau  = 
6 \pi S_{EW}\, i \oint_{|s|=m_\tau^2} {\frac{ds}{ m_\tau^2}}\,
\left( 1- {\frac{s}{ m_\tau^2}}\right)^2 \left[ \left( 
1 + 2 {\frac{s}{m_\tau^2}}\right) \Pi^{(0+1)}(s)
- 2 {\frac{s}{ m_\tau^2}}\, \Pi^{(0)}(s) \right] \, .
\label{OPErep}\end{equation}
Since $m_\tau^2\sim 3\ {\rm GeV}^2$, the latter expression is
amenable to evaluation using the OPE.
Keeping operators up to dimension $D=6$ in the OPE, and evaluating
the mass-independent ($D=0$) and mass-dependent ($D=2$) perturbative
terms to ${\cal O}(\alpha_s^3)$ and ${\cal O}(\alpha_s^2)$,
respectively, it is known that an excellent fit to hadronic
$\tau$ decay data is obtained using a value of $\alpha_s(m_\tau )$
compatible (after running) with that obtained directly at a
scale $M_Z$ (see Ref.~\cite{pich97}, and earlier references
cited therein).  Of relevance to the present paper is the
observation, made by numerous earlier authors, that, although the
$D=2$ $ud$ contributions to $R_\tau$ are tiny, the $D=2$
$us$ terms can alter the prediction for the $us$ contribution
to $R_\tau$ by $\sim 20\%$, for conventional values of $m_s$.
(See, for example, Ref.~\cite{tau4,tau6,davier96,chen97}.)  These
contributions, of course, play little role in determining $R_\tau$, owing
to the suppression by the factor of $|V_{us}/V_{ud}|^2$
relative to the non-strange current contributions.
As noted by Davier\cite{davier96}, however,
sufficiently precise $\tau$ decay data allow one to look explicitly
for this effect, and hence, in principle, determine $m_s$.

The most convenient way of implementing this
extraction of $m_s$ is to consider the contributions
to inclusive hadronic $\tau$ decay mediated separately by the
strangeness-changing ($us$), and strangeness-non-changing ($ud$)
weak currents, and take the difference of these contributions after
rescaling each by the inverse of the square of the corresponding
CKM matrix element.  The $D=0$ contributions then cancel in
the difference, leaving a leading $D=2$ contribution proportional
(to leading order in $m_u/m_s$) to $m_s^2$.  Numerical 
estimates show that this term is not only formally leading, but
also numerically leading at the scale $m_\tau$ fixed by
the radius of the circular contour in 
Eq.~(\ref{OPErep})\cite{tau4,tau6}.

The extraction of $m_s$, as performed by the
ALEPH Collaboration\cite{davier96,chen97},
using flavor breaking in hadronic $\tau$ decays, relies on the
expressions for the $D=2$ mass-dependent terms worked out by
Chetyrkin and Kwiatkowski\cite{tau6}.  Having determined the $D=2$
contributions to the correlators $\Pi^{(0+1),(0)}_{us;V,A}(q^2)$
and performed the contour integration indicated in Eq.~(\ref{OPErep}),
they obtain the result, to leading order in $m_u/m_s$, 
quoted in their Eq. (26).  This result implies, summing
``longitudinal'' ($J=0$) and ``transverse'' ($J=0+J=1$) contributions,
a $us$ current contribution to $R_\tau$ given by
\begin{equation}
\left[ R_\tau \right]^{D=2}_{us;V,A}=
-12\, |V_{us}|^2\, S_{EW}
{\frac{m^2_s(m_\tau )}{m^2_\tau}}\, \left[ 1 + {\frac{16}{3}}
a(m_\tau ) +11.03\, a(m_\tau )^2 \right]\ .
\label{ChKversion}\end{equation}
In Eq.~(\ref{ChKversion}), 
$m_s(m_\tau )$ is the running strange
quark mass at a scale $m_\tau$, and $a(\mu )\equiv\alpha_s(\mu )/\pi$,
with $\alpha_s(\mu )$ the standard $\overline{MS}$
running coupling at scale $\mu$.
The polynomial $P(a)=1+16a/3 +11.03 a^2$ appearing in 
Eq.~(\ref{OPErep}) is evidently rather well-converged to ${\cal O}(a^2)$,
allowing, once experimental data is sufficiently precise, a reliable
determination of $m_s(m_\tau )$.  However, as we will see below, the
coefficient $11.03$ multiplying $a^2$ in $P(a)$ is incorrect.  Not
only is the corrected value ($46.002$) such that the convergence
of $P(a)$ is not very compelling ($P(a)=1+.602 +.587+...$ at the
scale $m_\tau$) but, worse, this large value is generated mostly
by an enormous ${\cal O}(a^2)$ coefficient in the longitudinal
contribution.  The series for
the longitudinal contribution then
turns out to be such that each subsequent
term is larger than the previous, making it impossible to rely
on the OPE representation of this contribution, and hence impossible
to use the OPE representation of the D=2 terms to extract $m_s$.

We will, however, see that it is still possible to make some use
of the data on flavor-breaking in $\tau$ decay, since the 
spectral function of the badly behaved longitudinal contribution
is related to that occuring in existing sum rule analyses of
$m_s$ based on the strangeness-changing scalar channel.  Although
the spectral function has not been experimentally determined
in that channel, model versions 
exist\cite{jm,cps,cfnp}, and information on the low-energy portion of
it {\it is} available using the 
value of the scalar $K\pi$ form factor measured in $K_{\ell 3}$, together
with the Omnes representation of the form factor, which takes 
experimentally measured $K\pi$ phase shifts as input\cite{cfnp}.

The rest of this note is organized as follows.  In Section II we
present the corrected results for the longitudinal and 
transverse
$D=2$ $us$ contributions
to $R_\tau$ and demonstrate the non-convergence of the 
longitudinal series in $a(m_\tau)$.  In Section III we will see
how to reformulate the expression for the $us$ contribution to
$R_\tau$ in such a way that it involves only (1) a longitudinal
contribution which can be calculated directly from any
given (model-dependent or otherwise) version of the spectral
function for the strangeness-changing scalar channel,
(2) reasonably well-converged $D=2$ transverse contributions and (3) 
higher dimension longitudinal and transverse
contributions which are either quite accurately known
or small.  In Section IV we will then use existing models for
the scalar spectral function, together with recent $\tau$ decay
data, to perform a consistency check on the values of $m_s$
extracted using the two different analyses.

\section{The perturbative series for the D=2 contribution}
The $D=2$ $us$ contributions to $R_\tau$ follow immediately from
Eq.~(\ref{OPErep}) and the expressions for the longitudinal
and transverse pieces of the relevant vector and axial
vector current correlators to next-to-next-to-leading
order, as given by Chetyrkin and Kwiatkowski\cite{tau6}.
In what follows we will drop terms of order $m_u/m_s$ since,
as we will see below, it is possible to employ $\tau$ decay data
(once we are aware of the poor convergence of the longitudinal
contributions) only if we neglect such contributions.  Note
also that the same extremely poor convergence of the longitudinal
contributions also afflicts the $D=2$ $ud$ contributions; however,
in this case, the squared-mass factors are so tiny that this is
of no numerical significance in existing analyses of the 
isovector hadronic weak current mediated decays.  

The 
contributions to the correlators, in this approximation, are then
given by\cite{tau6}
\begin{equation}
\Pi^{(0)D=2}_{us;V,A}= {\frac{3}{2\pi^2}}{\frac{m_s(Q^2)}{Q^2}}
\left[ {\frac{1}{a(Q^2)}}-{\frac{5}{2}}-1.713804 a(q^2)\right]
+\cdots
\label{d2long}\end{equation}
and
\begin{equation}
\Pi^{(0+1)D=2}_{us;V,A}= -{\frac{3}{4\pi}}{\frac{
m_s(Q^2)}{Q^2}}\left[ 1+{\frac{7}{3}} a(Q^2)
+19.5831 a(Q^2)^2\right]
\label{d2trans}\end{equation}
where $a(Q^2) =\alpha_s(Q^2)/\pi$
and $m_s(Q^2)$ are the running coupling and running strange quark mass, both
at scale $\mu^2=Q^2=-s$, in the $\overline{MS}$ scheme.
In Eq.~(\ref{d2long}), $+\cdots$ stands for the second set of
terms of Eq. (16) of Ref.~\cite{tau6}, which terms do not survive the
contour integration involved in obtaining $R_\tau$.  Note that,
in Eq.~(\ref{d2trans}), the coefficient $19.5831$ includes
the contribution of
the three-loop graph containing two quark loops, which contribution
is also present in the corresponding $ud$ contribution to $R_\tau$
and hence cancels in forming the flavor-breaking difference
discussed below.  (The term which survives corresponds to the
replacement $19.5831\rightarrow 19.9332$.)  Note that there is a
typographical error in Eq.~(25) of Ref.~\cite{tau6}, where the
correct result $19.5831\sim 19.6$ has inadvertantly been
recorded as $17.6$.

It is now a straightforward matter to evaluate the $D=2$ $us$
contributions to $R_\tau$.  One first
expands $a(Q^2)$ in terms of $a(m_\tau^2)\equiv a$
and $m_s(Q^2)$ in terms of $a$ and
$m_s(m_\tau^2)\equiv \bar{m}_s$, the coefficients in these
expansions being polynomials in $\log{(Q^2/m_\tau^2)}$\cite{kniehl96}.
The remaining integrals are then elementary ones, involving products
of powers of $Q^2$ and powers of $\log{(Q^2/m_\tau^2)}$.
Performing these integrals, one finds, for the contributions
to $R_\tau$ corresponding to the terms written explicitly
in Eqs.~(\ref{d2long}) and (\ref{d2trans}) above,
\begin{eqnarray}
\left[ R_\tau^{(0)}\right]^{D=2}_{us;V,A}&=& -3|V_{us}|^2 S_{EW}\,
{\frac{\bar{m}_s^2}{m_\tau^2}}\left[ 1+
{\frac{28}{3}}a+109.9889 a^2\right] 
\label{longpoly} \\
\left[ R_\tau^{(0+1)}\right]^{D=2}_{us;V,A}&=& -9|V_{us}|^2 S_{EW}\,
{\frac{\bar{m}_s^2}{m_\tau^2}}\left[ 1+4a+24.6707 a^2\right] \ ,
\label{transpoly}\end{eqnarray}
whose sum yields, as claimed above,
\begin{equation}
\left[ R_\tau \right]^{D=2}_{us;V,A}= -12|V_{us}|^2 S_{EW}\,
{\frac{\bar{m}_s^2}{m_\tau^2}}\left[ 1+
{\frac{16}{3}}a+46.002 a^2\right] \ .
\end{equation}
(In the transverse contribution, $24.6707\rightarrow 25.0207$
when one takes into account the cancellation of the terms
proportional to $\alpha_s^2 m_s^2$ common to the $ud$ and $us$ correlators,
and associated with the graphs involving
a quark bubble on the internal gluon line.)
The revised numerical values quoted above have been confirmed by
Chetyrkin\cite{Chprivate}.

If one uses the value of $\alpha_s(m_\tau)$ determined from the ALEPH
analysis of non-strange hadronic $\tau$ decays\cite{ALEPHalphas} (this
choice being the most sensible since, in contrast to the global analysis, the
results are not affected numerically by the use of the erroneous
form of the $D=2$ contributions), one finds, for the polynomial
appearing in the Eq.~(\ref{longpoly}),
\begin{equation}
1+{\frac{2}{3}}8a+109.9889a^2=1+1.052+1.397\ .
\end{equation} 
It is quite clear that the series is simply too poorly converged 
to allow one to consider the coefficient of $m_s^2$ appearing
in the expression for $R_\tau$ reliably determined by the OPE
analysis.  As such, the conventional approach to obtaining $m_s$
from flavor breaking in hadronic $\tau$ decay is ruled out.

\section{A modified analysis for \mstitle}
Two observations make it possible to utilize flavor breaking in
hadronic $\tau$ decay to constrain $m_s$, in spite of the
non-convergence of the series for the longitudinal contribution
noted above.  The first is that, to the extent that one ignores
corrections of order $m_u/m_s$, the vector and axial vector
$D=2$ $us$ contributions are identical.  (Beyond leading order in $m_s$, this
statement is, however, no longer valid.)
One may then, as far
as the $D=2$ contributions are concerned, concentrate on, say,
the vector contributions.  This is of relevance because of
the second observation, namely 
that the vector current spectral function
is related to the spectral function of the strangeness-changing
scalar channel, about which some (though not complete) experimental
information is available.  Combining these two
observations, one sees that the problematic $D=2$ contributions
can be handled by replacing the full ($D=2,4,6,...$ and vector
plus axial vector) longitudinal
contributions with twice the full vector contribution, providing
one makes corrections reflecting the difference of the vector
and axial vector contributions.  These corrections then
(to leading order in $m_s$) begin at $D=4$.  The full vector
contribution is obtained, as described below, from the model
spectral functions employed in the strangeness-changing scalar
channel.

Let us define $R^{(J);D}_{ij;V,A}$ to be the vector ($V$) or
axial vector ($A$) flavor $ij=ud$ or $us$ contributions
to $R_\tau$ associated with
those operators of dimension $D$ in the OPE of
the longitudinal ($J=0$) or transverse ($J=0+1$) terms in Eq.~(\ref{OPErep})
{\it after scaling out the factors $|V_{ij}|^2\, S_{EW}$}, and
$R_{ij}$ to be the corresponding sum, for given $ij$, over $D$, $V+A$
and $J=0,0+1$.  As noted above, to be able to effectively employ the model
spectral functions from the scalar channel as input it is necessary
to drop corrections of ${\cal O}(m_u/m_s)$ to $R^{(0);D=2}_{us;V,A}$.
For consistency one must then also drop other terms proportional
to the light quark $u,d$ masses.  For $D=2$, this includes all of
$R^{(J);D=2}_{ud;V,A}$ apart from the ``quark bubble'' transverse
contribution proportional to $\alpha_s^2 m_s^2$ mentioned above,
which simply cancels against the corresponding term in
$R^{(0);D=2}_{us;V,A}$.
For $D=4$, since the leading contributions are proportional 
to $<m_s \bar{\ell}\ell >$, $\ell =u,d$, 
and $m_s^4$, this means dropping also
terms proportional to $<m_\ell \bar{\ell}\ell>$, and terms fourth
order in the quark masses, other than those proportional to $m_s^4$.
This removes all contributions to $R^{(J);D=4}_{ud;V,A}$ apart
from those terms in the transverse
contribution proportional to the gluon condensate, which terms 
cancel in forming the difference of $ud$ and $us$ contributions.
Note also that, to the order so far computed, the $D=6$ contributions
are pure transverse.
With the set of approximations just discussed,
one thus obtains,
\begin{eqnarray}
R_{ud}-R_{us}&=& - \left[ 2\sum_{D}\, R^{(0);D}_{us;V}\right] 
- \left[ R^{(0);D=4}_{us;A} - R^{(0);D=4}_{us;V}\right]
-\left[ R^{(0+1);D=2}_{us;V}+R^{(0+1);D=2}_{us;A}\right]^\prime \nonumber\\
&&\ \ +\sum_{D=4,6,\cdots}\,
\left[ R^{(0+1);D}_{ud;V} +R^{(0+1);D}_{ud;A}
-R^{(0+1);D}_{us;V}-R^{(0+1);D}_{us;A}\right] \ ,
\label{approx}\end{eqnarray}
where terms with $D>6$ have been dropped.  
The first term on the RHS of Eq.~(\ref{approx}) is to be obtained using
the input model spectral function obtained from the scalar channel.
The remaining terms will be obtained from the OPE representation.
The prime on the third 
term in Eq.~(\ref{approx}) reminds us that the ``strange quark bubble''
contribution cancelled by the corresponding contribution to the
$ud$ correlator has to be removed in evaluating this term.
The $ud$ contributions in the last term serve only to remove
all $D=4$ contributions proportional to the gluon condensate.
As we will see below, all of the remaining terms are either very
well, or relatively well, known.  Eq.~(\ref{approx}) thus isolates
the problems of the $D=2$ part of the 
longitudinal term in the OPE representation in such a way that,
to the extent one can obtain a good phenomenological representation
of the scalar spectral function, one can avoid the poorly converging
OPE representation and instead evaluate this contribution phenomenologically.
While, at present, it is not possible to do this in a completely
satisfactory manner (see the discussion on the existing constraints
on the scalar spectral function below), the situation is amenable to
future improvement.

Let us turn then to the evaluation of the remaining terms in 
Eq.~(\ref{approx}).  We stress again, that in writing down all
the expressions below, corrections proportional to the light quark
masses have, for consistency's sake, been systematically dropped, 
with the exception of one term for which there is a numerical
enhancement, as discussed below.
Since the expressions for the relevant contributions to the correlators
are well-known in the literature\cite{tau4,tau6},
we present only the OPE representations after the 
relevant contour integrations
have been performed.
We then find,
\begin{eqnarray}
-&& \left[ R^{(0);D=4}_{us;A} - R^{(0);D=4}_{us;V}\right]=
{\frac{24\pi^2}{m_\tau^4}}\left[ -2\bar{m}_s<\bar{\ell }\ell>\right]
+12\left[ {\frac{12}{7a}}+{\frac{5}{7}}\right]
\left({\frac{ \bar{m}_s}{m_\tau}}\right)^4
\left({\frac{ m_\ell}{m_s}}\right)
\nonumber \\
-&&\left[ R^{(0+1);D=2}_{us;V}+R^{(0+1);D=2}_{us;A}\right]^\prime
=-18\left[ {\frac{\bar{m}_s}{m_\tau}}\right]^2\left[ 1+4a+25.0207a^2
+f_3a^3\right]
\nonumber \\
&&\left[ R^{(0+1);D=4}_{ud;V} +R^{(0+1);D=4}_{ud;A}
-R^{(0+1);D=4}_{us;V}-R^{(0+1);D=4}_{us;A}\right]=
{\frac{81\pi^2a^2}{m_\tau^4}}\left[ -2\bar{m}_s<\bar{s }s>\right]
\nonumber\\
&&\qquad\qquad\qquad\qquad\qquad\qquad\qquad
-54\left({\frac{ \bar{m}_s}{m_\tau}}\right)^4 \nonumber\\
&&\left[ R^{(0+1);D=6}_{ud;V} +R^{(0+1);D=6}_{ud;A}
-R^{(0+1);D=6}_{us;V}-R^{(0+1);D=6}_{us;A}\right]\simeq 0\ .
\label{opecorr}\end{eqnarray}

A few comments are in order concerning Eqs.~(\ref{opecorr}).  First,
the leading term fourth order in the quark masses in the first
expression, though formally an ${\cal O}(m_\ell /m_s)$ correction
to the $m_s^4$ contribution appearing in the third expression,
has a numerical enhancement due to the presence of a factor $1/a$
in the coefficient multiplying it, which enhancement largely undoes
the $m_\ell /m_s$ suppression.  This term has, therefore, been
kept, though it turns out that, in fact, both terms
are numerically tiny, and play a negligible role
in the final analysis.  The presence of a $1/a$ factor in the coefficient
multiplying $m_s^3m_\ell$ is due to the well-known fact that the quark
condensate terms obtained from the standard Wick's theorem reduction 
must be modified to absorb additional long-distance mass logarithm terms.
The modified condensates are
then no longer scale invariant.  Defining new scale invariant versions
(see Ref.~\cite{tau4} for details) one finds that, rewriting
the $D=4$ contributions in terms of these condensates, such inverse
power dependence on $a$ shows up in the modified fourth order
mass terms.  The quark condensates appearing in Eq.~(\ref{opecorr})
are thus those defined in the Appendix of Refs.~\cite{d4con,tau4}.
Second, in the $D=2$ transverse terms of the second line, the
coefficient $f_3$ of $a^3$ has not, as yet, been computed.
Because of the slow convergence of the ${\cal O}(a, a^2)$
terms in the series, we have, however, made an estimate of $f_3$.  This
is done by first estimating the corresponding coefficient in the
series for the $D=2$ contribution to the correlator using
the method of Ref.~\cite{cks}, and then integrating the result
around the circular contour.  The proceedure of Ref.~\cite{cks}
is known to produce values for the ${\cal O}(a^3)$
coefficient accurate to $\pm 25$
in the case of those mass-dependent observables for which the
${\cal O}(a^3)$ terms have been computed\cite{cks}.  We find
$f_3=217.8$ from the ``fastest apparent convergence'' version
of the method, and $f_3=219.6$ for the ``principle of minimal
sensitivity version''.  As a conservative estimate of the
effect of higher order terms we thus take $f_3=220\pm 220$.
At present the uncertainties in $\bar{m}_s$ that result
from that in $f_3$ are much smaller than those associated
with the errors in the experimental input.  Finally, the reason
for setting the $D=6$ (purely transverse) contributions to
zero needs explanation.  It is straightforward to read off
the explicit forms of these contributions from Eq. (3.12)
of Ref.~\cite{tau4}.  As discussed in that reference, however,
the full set of $D=6$ condensates is not known empirically,
and it is customary to make an estimate based on a rescaled
version of the vacuum saturation hypothesis (Eq. (3.13) of
Ref.~\cite{tau4}).  This estimate has relatively large 
errors, but can be checked by employing the method of 
spectral moments\cite{tau5,pich97}, which allows
one to make an empirical
extraction of (at least the $ud$) $D=6$ contributions using
the measured spectral data.  This extraction has been
performed by the ALEPH collaboration\cite{ALEPHalphas,ALEPHspec}, who obtain, 
for the sum of vector and
axial vector $ud$ $D=6$ contributions, a value seven times smaller
than the central value obtained in Ref.~\cite{tau4}, but
compatible with it, within the estimated theoretical errors.  With this
empirical information as input, and the realization that the
difference of the $ud$ and $us$ $D=6$ contributions should
be further suppressed by the approximate $SU(3)_F$ flavor 
symmetry, the numerical value of the combined $D=6$ contribution,
though not well-determined, becomes completely negligible
numerically, and hence is neglected in the analysis which follows.

\section{A modified extraction of \mstitle}
In what follows we detail the input required to numerically evaluate
the contributions listed in the previous section.  We employ, from
the various ALEPH fits for $a=\alpha_s(m_\tau )/\pi$, that obtained from
the analysis of the non-strange modes alone, since the slow
convergence of the longitudinal contributions in the strange case
make that extraction unreliable.  In quoting values for $m_s$,
we will also run the value extracted at scale $m_\tau$ down to
a scale $1\ {\rm GeV}^2$, using the recently-determined
four-loop $\beta$\cite{beta4} and $\gamma$\cite{gamma4} functions,
in order to provide comparisons to other results for 
$m_s$\cite{jm,cps,cfnp,narison95,km97}
all of which are quoted at that lower scale.
Other necessary inputs, apart from the longitudinal vector contribution, to
be discussed in detail below, are the light quark mass ratio
$m_s/\hat{m}=24.4\pm 1.5$\cite{leutwyler96} (where $\hat{m}=(m_u+m_d)/2$)
and
\begin{equation}
-2m_s <\bar{s} s>=\left( {\frac{m_s}{\hat{m}}}\right) \left(
{\frac{<\bar{s}s>}{<\bar{u}u>}}\right) 
\left( -2\hat{m}<\bar{u}u>\right)
= \left( {\frac{m_s}{\hat{m}}}\right) \left(
{\frac{<\bar{s}s>}{<\bar{u}u>}}\right) f_\pi^2m_\pi^2\ ,
\label{d4diffval}\end{equation}
where, following Refs.~\cite{jm,cps}, we take the ratio of strange
to light quark condensates
to lie between $0.7$ and $1$.

The crucial piece of input, however, which can, in principle
at least, allow us to evade the problems with the longitudinal $D=2$
terms, is the full longitudinal vector contribution,
$\sum_D \left[ R^{(0);D}_{us;V}\right]$.  This can be obtained
straightforwardly
from any model (or empirical determination) of the spectral function
of the strangeness-changing scalar channel since the longitudinal $us$
vector current spectral function is simply $1/s^2$ times that
of the scalar channel.  The latter spectral function begins at
$K\pi$ threshold, the next open channel being $K\pi\pi\pi$
(experimentally, however, the $K\pi$ channel is purely elastic 
below $(1.3\ {\rm GeV})^2$\cite{kpiexp}).  Given
that the $K_0^*(1430)$ has a branching ratio of $93\pm 4\pm 9\, \%$ to 
$K\pi$\cite{PDG96}, it is safe to assume that the $K\pi$ mode
dominates the scalar spectral function out to $s\sim 2\ {\rm GeV}^2$.
This observation is particularly relevant because the $K\pi$ contribution
to the spectral function is determined by the timelike $K\pi$
scalar form factor, which satisfies an Omnes representation\cite{jm}.
Using as input the experimental value of the form factor determined
in $K_{\ell 3}$, together with the measured $K\pi$ phase shifts,
it is thus, in principle, possible to determine the $K\pi$ contribution to the
physical scalar spectral function\cite{jm,cfnp}.  
In practice, since the $K\pi$ phases have only been measured out to
$s=(1.7\ {\rm GeV})^2$, certain assumptions are required about 
the behavior of the phase beyond this point.  In Ref.~\cite{cfnp}
it has been assumed that the asymptotic value, $\pi$, of the phase (associated
with the known asymptotic $1/s$-dependence of the scalar form factor)
is achieved for all $s>(1.7\ {\rm GeV})^2$.  At present, neither
theoretical nor experimental checks of this assumption exist, though
the experimental phase does approach the asymptotic value 
as $s\rightarrow (1.7\ {\rm GeV})^2$.  One may also check that, if for
example the phase were to make an excursion into the third quadrant as
one passed through the $K_0^*(1950)$ region, and then return to
$\pi$, the longitudinal spectral integral relevant to $\tau$
decay would be increased by less than $\sim 10-15\%$, producing
a decrease of $m_s$ of less than $\sim 2\%$.
Were
the onset of the asymptotic behavior to occur significantly beyond
this point, however, the scalar form factor obtained in Ref.~\cite{cfnp}
could be significantly altered, even at low $s$.  This difficulty
is, of course, potentially surmountable in future.
Lacking any
additional information at present, and in light of the discussion above,
we will tentatively accept
the assumption of Ref.~\cite{cfnp}, and hence consider the $K\pi$
portion of the spectral function, and hence the full
scalar spectral function below $\sim 2\ {\rm GeV}^2$, to be
as determined there.  

Even if this assumption is correct,
however, this leaves the problem 
that one expects additional channels to become 
increasingly important as $s$ is increased beyond the
location of the $K_0^*(1430)$.  Since the $K_0^*(1950)$
$K\pi$ branching fraction is only $52\pm 14\, \%$\cite{PDG96}, it
is likely that employing only the $K\pi$ contribution to the full
spectral function will become a questionable approximation significantly
before one has reached the end of the range of the spectral integral
relevant to $\tau$ decay.  This likelihood is reinforced by the
observation that, attempting to perform a standard Borel transformed
QCD sum rule analysis of the scalar channel using only the $K\pi$
portion of the spectral function determined in Ref.~\cite{cfnp},
one finds no stability plateau for the ``extracted'' strange quark mass
in the range $s\sim 3\ {\rm GeV}^2$\cite{jamin97}.  Since
\begin{equation}
-2\, \sum_D \left[ R^{(0);D}_{us;V}\right] = 12\pi^2 S_{EW}\,
\int_0^{m_\tau^2}\, {\frac{ds}{m_\tau^2}}\left( 1-2{\frac{s}{m_\tau^2}}
\right)^2\left( 4 {\frac{\rho_{us}^{s}(s)}{m_\tau^2s}}
\right)\label{model}\end{equation}
where $\rho_{us}^s(s)$ is the spectral function of the scalar
strange channel, such an underestimate of the spectral function
corresponds to an {\it underestimate} of the full longitudinal contribution
to $\tau$ decay, and hence (because of the overall sign ($+$) with which
the contribution in Eq.~(\ref{model}) enters $R_{ud}-R_{us}$),
to an {\it overestimate} of $m_s$.  While the weight function
entering the spectral integral in Eq.~(\ref{model}) has a double
zero at $s=m_\tau^2$, this does not help the situation as much
as one might hope:  although the maximum of the weight function occurs
for $s/m_\tau^2 =1/3$, the weight has decreased by only a factor of $2$
from its maximum value for $s/m_\tau^2 =2/3$.  At present there is, thus,
no satisfactory extension of the full spectral function beyond
$s\sim 2 \ {\rm GeV}^2$, and this limits our ability to extract
$m_s$ from the hadronic $\tau$ decay data.  Note, however, that,
while neglect of the non-$K\pi$ portion of the spectral function
leads to an overestimate of $m_s$ when one employs the $\tau$
decay analysis, it leads to an {\it underestimate} of $m_s$
when one employs the direct QCD sum rule analysis of the 
scalar channel.  The extent to which the two different extractions
of $m_s$ (both of which depend on the longitudinal spectral function)
agree thus provides some {\it post facto} information of the
degree of reliability of the model spectral function.

Using the $K\pi$ portion of the spectral function as determined in
Ref.~\cite{cfnp}, we then find 
\begin{equation}
-2\, \sum_D \left[ R^{(0);D}_{us;V}\right] = .055\ .
\label{longval}\end{equation}
One can only guess at the error on this number at present.  Varying
the parameters of the fit to the experimental phases as described
in Ref.~\cite{cfnp} one finds a variation of $\pm .0012$, while,
as explained above, the uncertainty associated with the unknown
behavior of the $K\pi$ phase above $(1.7\ {\rm GeV})^2$ could
easily produce uncertainties of order $10-15\%$ or more.  By way
of contrast, the model spectral function employed in Refs.~\cite{jm,cps}
produces a value $.090$.  Note, however, that, as pointed out in
Ref.~\cite{cfnp}, the behavior of this spectral function
below $s\sim 2\ {\rm GeV}^2$ is incompatible with the Omnes
representation and known $K\pi$ phases, and hence is unphysical.

For the experimental input to the analysis we employ the following.
First, for the strange and non-strange current contributions to $R_\tau$
in Eq.~(\ref{one}), we employ the latest ALEPH
results\cite{davier96,chen97,ALEPHspec,hocker}
\begin{eqnarray}
R_{\tau ;ud}&=&3.493\pm .026\nonumber \\
R_{\tau ;us}&=&0.155\pm .008\ .\label{ALEPHr}
\end{eqnarray}
These are compatible with
those of CLEO\cite{CLEOstrange,weinstein}; 
the latter collaboration, however, has not yet released
an official number for the strange branching fraction\cite{weinstein}.
For the CKM matrix elements, we take the value of $V_{us}$ from
$K_{e3}$ data, where the extraction is under best theoretical
control, and $V_{ud}$ from three-family unitarity (there exist, for
example, effective isospin-breaking electromagnetic contact
interactions, which involve no explicit photons in the effective
hadronic theory, and which are, therefore, not taken into
account in treatments of super-allowed $\beta$ decays; it
is not clear how to determine a probable error associated with
this neglect).  With this input, and the electroweak correction
from Ref.~\cite{ms88}, we obtain
\begin{equation}
\left[ R_{ud}-R_{us}\right]_{expt} = 0.446\pm 0.169\ .
\label{exptdiff}\end{equation}
The cancellation present in Eq.~(\ref{exptdiff}), of course,
magnifies the errors.  For example, if we employed the older
ALEPH value $R_{\tau ;us}=0.156$ the central value would be
shifted from $0.446$ to $0.426$.  The overall error quoted
is dominated by that on $R_{us}$.

From Eqs.~(\ref{approx}), (\ref{opecorr}),
(\ref{d4diffval}), (\ref{longval}) and (\ref{exptdiff}), one obtains
the following upper bound for $m_s$
\begin{equation}
m_s(1\ {\rm GeV}^2)\, < \, 220^{+56}_{-77}\pm 33\ {\rm MeV}\, ,
\label{tauresult}\end{equation}
where the first set of (asymmetric) errors is associated with the
experimental uncertainties in Eq.~(\ref{exptdiff}), and the second with
the estimate above of the uncertainties due to higher order contributions
in the perturbative series for the $D=2$ transverse terms.

In quoting errors in Eq.(\ref{tauresult}) we have refrained from including 
those due to possible shortcomings in the model longitudinal spectral
function (associated with (1) the unknown high energy behavior of
the $K\pi$ phase and (2) neglected $K\pi\pi\pi , \cdots$ contributions
to the spectral function).  We remind the reader that these corrections,
if made, would produce a value {\it lower} 
than the bound shown in Eq.~(\ref{tauresult}),
though it is not possible to make a sensible estimate of the size of
this correction at present.  In contrast, the same set of corrections
would serve to {\it raise} the value
\begin{equation}
125\ {\rm MeV}\, < \, m_s(1\ {\rm GeV}^2)\, < \, 160\ {\rm MeV}\, ,
\end{equation}
obtained from the analysis of the scalar sum rule in Ref.~\cite{cfnp}.
The true result for $m_s$ will, in general, lie between the values
obtained from the $\tau$ decay analysis and the scalar sum rule,
so long as one uses only the $K\pi$ portion of the longitudinal
spectral function in both analyses.
At present, the two results, owing to the large errors, especially in
Eq.~(\ref{tauresult}), are compatible, but one would like to see
the errors in both reduced.

To conclude, we stress that flavor breaking in hadronic $\tau$ decays 
can still be employed as a means of extracting $m_s$, despite
the poor convergence of the perturbative series for the $D=2$
longitudinal terms, provided empirical input
can be obtained for the longitudinal spectral function.  As pointed
out in Refs.~\cite{kmzpc92,sfk}, it is possible to separate the
longitudinal and transverse components experimentally by analyzing
the dependence of the cross-section on the angle, $\beta$, between
the direction of the $K$ and the laboratory, as seen from the
hadronic rest frame.  One can thus, using $\tau$ decay data alone,
obtain an estimate for $m_s$, in the approximation that one
neglects the higher multiplicity ($K\pi\pi\pi ,\cdots$) contributions
to the longitudinal spectral function.  Since measuring the
$K\pi$ component of the longitudinal spectral function would
simultaneously determine the $K\pi$ component of the spectral
function for the scalar channel, a comparison of the $m_s$ values
extracted using $\tau$ decay and the scalar channel sum rule
would set upper and lower bounds for $m_s$, and provide an
explicit check of the reliability of the the approximation 
of neglecting higher multiplicity intermediate states.  The
experimental determination of the $K\pi$ component of the
longitudinal spectral function is, of course, also of interest
in that it would provide implicit information on the high
energy behavior of the $K\pi$ phase.
\acknowledgements
The author would like to thank K. Chetyrkin for re-checking the
earlier calculations of Ref.\cite{tau6} and confirming the
corrected values of the coefficients appearing in the $D=2$ terms.
Thanks also to Andreas H\"ocker of ALEPH and Alan Weinstein of 
CLEO for clarifying the status of results from their respective
collaborations.
Both the ongoing support of the Natural Sciences and
Engineering Research Council of Canada, and the hospitality of the
Special Research Centre for the Subatomic Structure of Matter at the
University of Adelaide, where most of this work was originally 
performed are also
gratefully acknowledged.

\end{document}